\documentclass[preprint,superscriptaddress]{revtex4}

\usepackage{graphicx}
\usepackage{amsmath}
\usepackage{amssymb}
\usepackage{epsfig}
\usepackage{wasysym}
\usepackage{bbm}
\usepackage[colorlinks,citecolor=blue]{hyperref}
\usepackage{tabularx}

\usepackage{color}

\renewcommand{\v}[1]{{\boldsymbol{#1}}}

\newcommand{\Ref}[1]{Ref.~\cite{#1}}
\newcommand{\Eq}[1]{equation~(\ref{#1})}
\newcommand{\Fig}[1]{Fig.~\ref{#1}}

\newcommand{\Rmnum}[1]{\expandafter\@slowromancap\romannumeral #1@}

\newcommand{\be}{\begin{eqnarray}}
\newcommand{\ee}{\end{eqnarray}}
\newcommand{\beq}{\begin{equation}}
\newcommand{\eeq}{\end{equation}}
\newcommand{\bpm}{\begin{pmatrix}}
\newcommand{\epm}{\end{pmatrix}}
\newcommand{\bal}{\begin{aligned}}
\newcommand{\eal}{\end{aligned}}

\newcommand{\ra}{\rightarrow}

\newcommand{\w}{{\omega}}

\newcommand \ti[1]{}

\begin{document}

\title{Topological versus Landau-like phase transitions}
\author{Lokman Tsui}
\affiliation{Department of Physics, University of California, Berkeley, CA 94720, USA.}
\author{Fa Wang}
\affiliation{International Center for Quantum Materials, School of Physics, Peking University, Beijing 100871, China.}
\affiliation{Collaborative Innovation Center of Quantum Matter, Beijing 100871, China.}
\author{Dung-Hai Lee}
\affiliation{Department of Physics, University of California, Berkeley, CA 94720, USA.}
\affiliation{Materials Sciences Division, Lawrence Berkeley National Laboratory, Berkeley, CA 94720, USA.}

\begin{abstract}
The study of continuous phase transitions triggered by spontaneous symmetry breaking has brought 
new concepts that revolutionized the way we understand many-body systems.  Recently, through the discovery of symmetry protected topological phases, it is realized that quantum phase transition between states with the same symmetry but different topology can also occur continuously. Here we ask ``what distinguishes these two types of phase transitions''.  
\end{abstract}

\maketitle

\begin{figure*}[h]
\includegraphics[scale=.325]{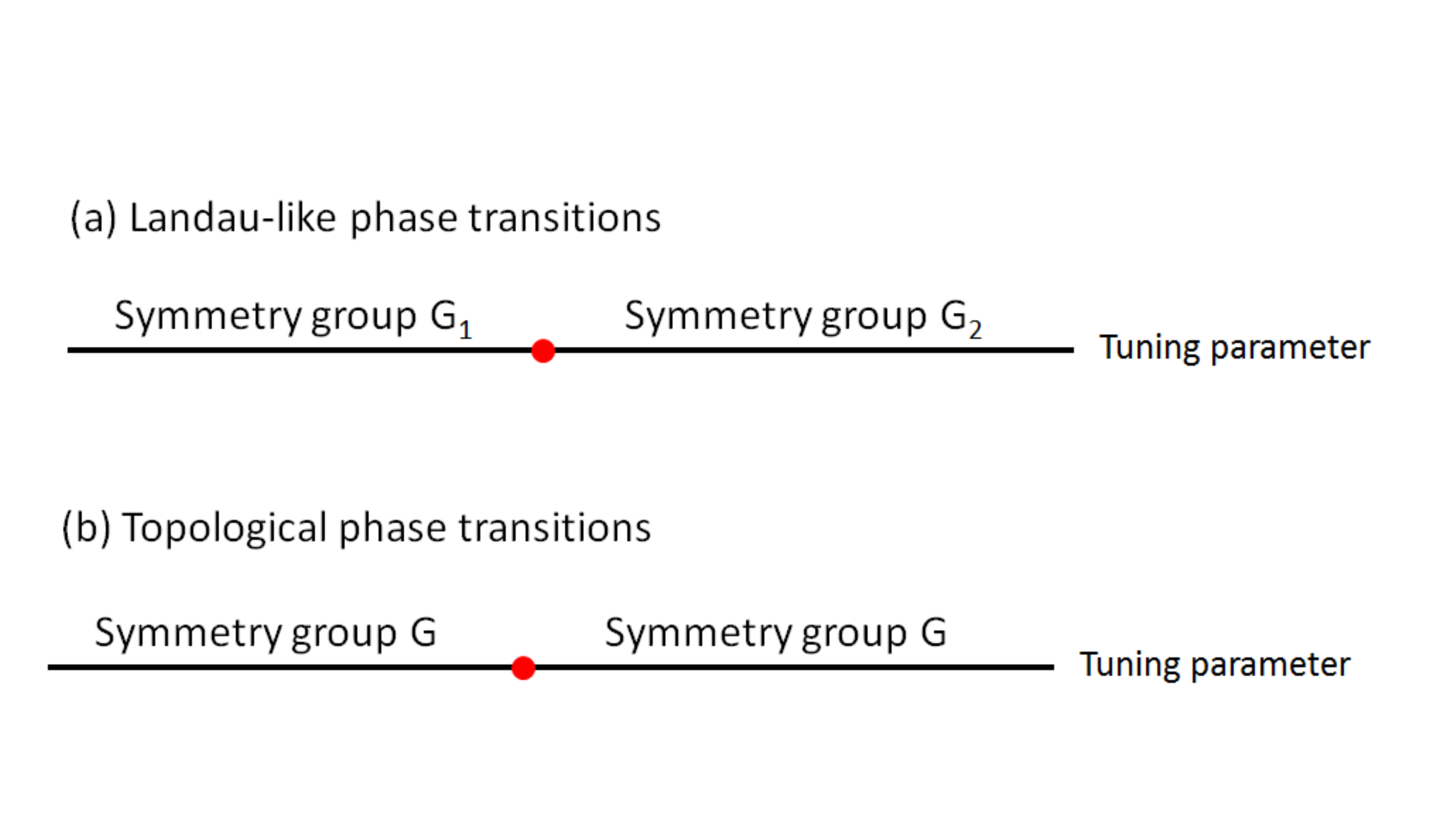}
\caption{Two types of continuous phase transitions. (a) Landau-like transitions. (b) Topological phase transitions. }
\label{transitions}
\end{figure*}

Continuous phase transitions triggered by spontaneous symmetry breaking (which we referred to as ``Landau-like phase transitions'') is an ubiquitous nature phenomenon. The ideas, e.g., the renormalization group and  conformal symmetry, developed for the understanding of them have impacted all areas of physics. In fact renormalizable continuum quantum field theories can be understood as conformal invariant field theories plus perturbations.
A Landau-like phase transition occurs between two phases with different symmetry groups $G_1$ and $G_2$  (see \Fig{transitions}(a)).According to Landau, if $G_1$ is a subgroup of $G_2$ (or vice versa) a continuous phase transition can occur generically. On the other hand if $G_1=G_2$ or if there is no subgroup relationship, the transition should be generically first order.\\

Last five years witnessed a fast progress in the understanding of a new type of quantum disordered states, namely, symmetry protected topological states (SPTs)\cite{Chen2011}. 
These states exhibit a full gap in the excitation spectrum when there is no boundary and do not break any Hamiltonian symmetry. 
Nonetheless these states are grouped into different ``topological classes'' such that interclass transitions must be accompanied by a phase transition with the closure of the energy gap. The purpose of this paper to understand the difference (if any) between the Landau type and this new kind of ``topological phase transitions'' where the two phases have the same symmetry (see \Fig{transitions}(b)).\\

\begin{figure*}[h]
\includegraphics[scale=.3]{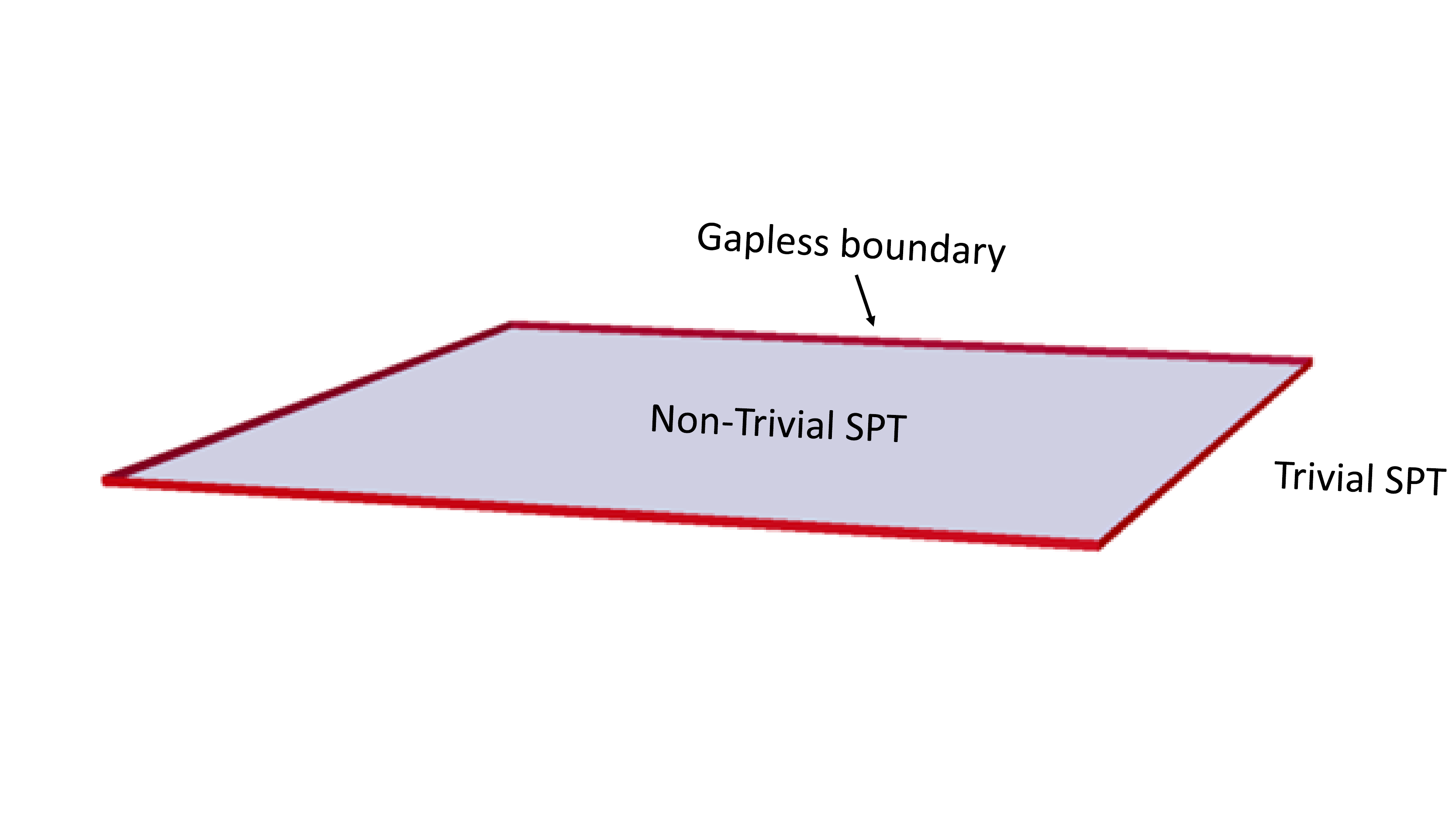}
\caption{The gapless boundary excitations of a non-trivial SPT can be viewed as due to a coordinate-tuned topological phase transition.}
\label{boundary}
\end{figure*}

A hallmark of  non-trivial SPTs is the existence of gapless boundary excitations. These excitations can intuitively be understood as a ``coordinate tuned topological phase transition''. Indeed, as an observer travels from inside a non-trivial topological phase to outside (the trivial vacuum) a gap closing transition  occurs at the boundary (for a two dimensional example see \Fig{boundary}).  On the other hand one can imagine tuning a parameter of the Hamiltonian so that a {\it bulk} phase transition occurs between the same non-trivial and trivial phases (\Fig{transitions}(b)). Viewing the boundary gapless states as those occuring at the critical point of a coordinate tuned SPT phase transition suggests there is an intimate relationship between the excitations of the bulk critical point and the boundary excitations of the non-trivial SPT phase. In \Ref{Chen2013} it is conjectured that the former are just a delocalized (or dynamically percolated) version of the latter.

\begin{figure*}[h]
\includegraphics[scale=.8,angle =0]{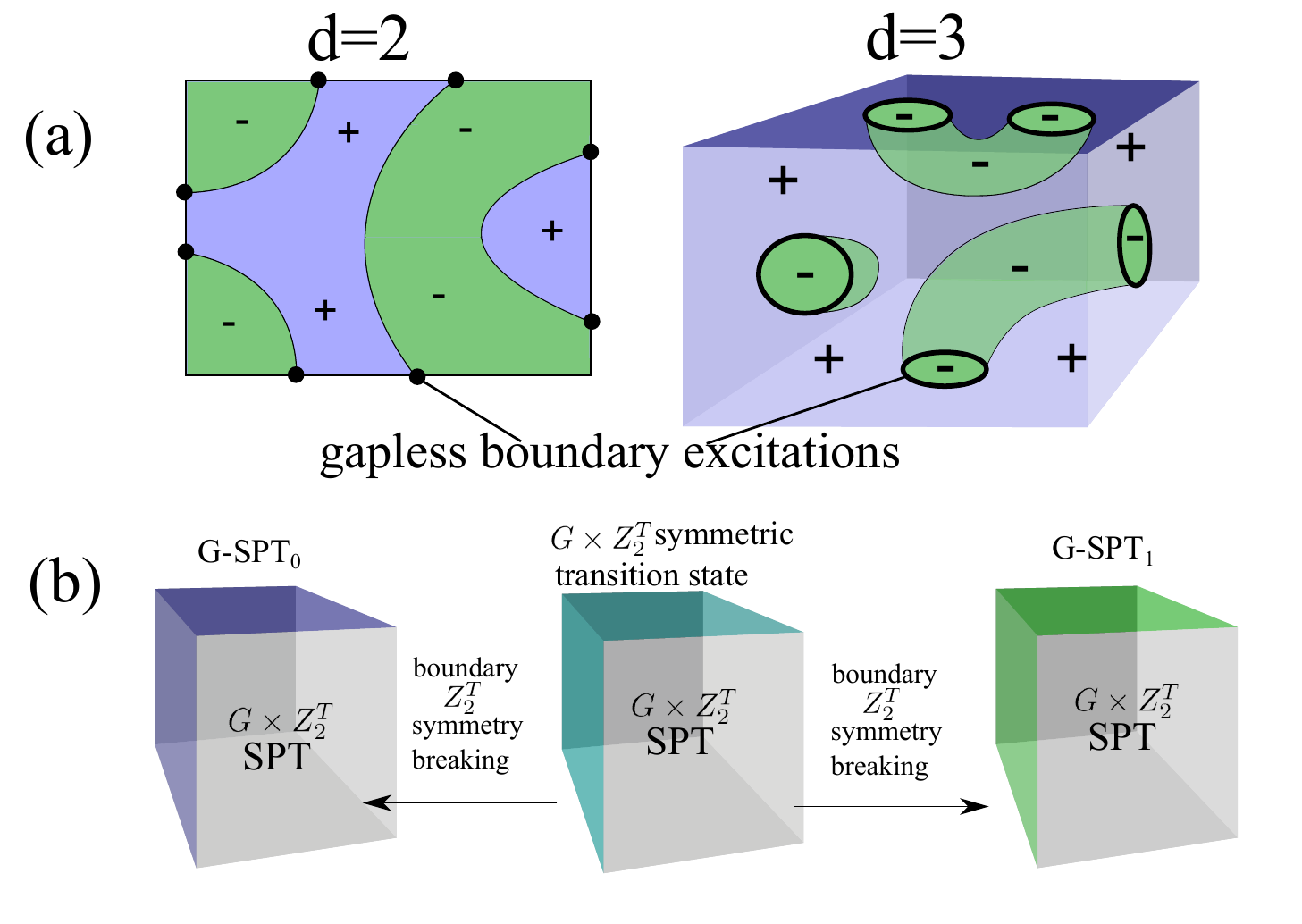}
\caption{
(a) The higher dimensional $G\times Z_2^T$ symmetric SPT constructed from proliferation of Ising domain walls each decorated with the non-trivial $G$-symmetric SPT whose transition into the trivial state is under study. Left panel: a two dimensional example, right panel: a three dimensional example.(b) The SPT phase transition on the {\it boundary} induced by a boundary $Z_2^T$ symmetry breaking field.}
\label{prolif}
\end{figure*}

As a sanity check we recall that when $G=SO(3)$ and when the space dimension is one, the following model exhibits a continuous phase transition between the trivial and non-trivial SPT phases. (We recall that in one dimension there are just two inequivalent $SO(3)$ protected SPT phases\cite{Chen2010,Chen2011}.) 
\be
H=(J+\delta)\sum_{i=odd} \v S_{i}\cdot\v S_{2i+1}+(J-\delta)\sum_{i=even} \v S_{i}\cdot\v S_{i+1},\ee where $\v S$ is a spin-1/2 operator and $J>0$. Each unit cell is composed of two sites with total spin equal to integers. Hence the degrees of freedom in each unit cell transform according to linear representations of $SO(3)$. The trivial and non-trivial phase transition is tuned by varying $\delta$ from negative to positive. The critical theory at $\delta=0$ is the famous Heisenberg chain, which is gapless. Moreover the Heisenberg chain is known to possess ``spinon'' (spin 1/2) excitations. Since the non-trivial SPT state have spin 1/2 excitations at both ends, it is consistent with the notion that the critical state possesses delocalized boundary excitations. The spinon excitation in the Heisenberg chain is a paradigmatic example of ``quantum number fractionalization''. A new understanding enabled by the above discussions is that the Heisenberg chain is the critical theory an SPT phase transition and the fractionalized excitations are derived from the boundary states of the non-trivial SPT. \\       
The conjecture made in \Ref{Chen2013} is verified (for general spatial dimension) in \Ref{Lokman2015} for a large class\footnote{The readers are referred to \Ref{Lokman2015} to find out what types of SPT phase transitions are covered by the theory.} of phase transitions between {\it bosonic} SPTs. The idea is to view the critical state between two d-dimensional $G$-symmetric SPTs as the boundary state of a SPT living in d+1 dimensions. Moreover, this d+1 dimensional SPT is protected by an enlarged symmetry group $G\times Z_2^T$ and its basic degrees of freedom on each lattice site are those of the $G$-symmetric SPT tensored with an Ising-like variable. The $Z_2^T$ transformation reverses the sign of the Ising variable and complex conjugates the Hamiltonian (and wavefunctions). The d+1  dimensional SPT is constructed as a quantum state arising from ``proliferated'' Ising domain wall with each wall ``decorated'' with a non-trivial $G$-symmetric SPT (see \Fig{prolif}(a))\footnote{When $Z_2^T$ is replaced by $Z_2$ this kind of decorated domain wall picture for SPT has been discussed in \Ref{Chen2014}. However in that case the only type of lower dimensional SPT that can be used to decorate the domain wall must satisfy the square of its wavefunction describes a trivial SPT.}. The readers are referred to \Ref{Lokman2015} for the wavefunction 
corresponding to such a d+1 dimensional SPT. In this ``holographic'' picture the phase transition in question entirely occurs on the boundary. It is induced by turning on a {\it boundary} symmetry breaking breaking field ($h$) which breaks the $Z_2^T$ symmetry (see \Fig{prolif}(b)).\\

In the above picture the critical excitations are the result of the interactions of the Ising domain wall with the boundry of the d+1 dimensional system. Because these intersections are themselves the boundary of the domain wall, they are infested with the gapless boundary excitations of the SPT residing on the domain wall. As the domain walls fluctuate these gapless excitations can move, pair annihilate, and pair create. In this way we understand the d-dimensional critical state as the dynamically percolated boundary excitations of the non-trivial SPT in question. \\     
  
In general the $Z_2^T$ or the $G$ symmetry of the d+1 dimensional SPT in the holographic picture can be spontaneously broken on the boundary without collapsing the bulk energy gap. In the former case the phase transition (between the two $G$ symmetric SPT) on the boundary proceeds as a first order phase transition. In the latter case there is an intermediate $G$ symmetry breaking phase separating the two $G$-symmetric SPTs. The critical points separating this intermediate phase from the two $G$-symmetric phases are Landau-like phase transitions. For details the readers are referred to \Ref{Lokman2015}. In the rest of the paper we shall focus on the most interesting scenario, namely continuous phase transitions.\\
    
Since continuous Landau-like phase transitions are best understood in 1+1 D (thanks to  conformal field theory), in the rest of the paper we focus on 1+1 D {\it bosonic} SPT phase transitions where conformal invariance is exhibited at the critical point\footnote{Among other things this requires the dynamical critical exponent to be 1}. Since rational conformal field theories are well classified in 1+1 D, our question becomes ``which subset of the rational conformal field theories can be realized as topological phase transitions''. \\

In \Ref{Chen2010} it is shown that the boundary degrees of freedom of a (non-trivial) 1-dimensional SPT carry  projective representations of the protection symmetry group. Since the critical state possesses delocalized boundary excitations, it naturally leads one to suspect that at the critical point between the trivial and non-trivial SPTs there are bulk excitations which transform projectively under the protection symmetry group. Before proceeding further in the following we prove that the projective representations of any group cannot be one dimensional. \\

{\bf Theorem:} the projective representations of any group $G$ cannot be one dimensional.\\
{\bf Proof:} Let $g_1,g_2$ be any two elements of $G$. Let $R(g_1)$ and $R(g_2)$ be their respective projective representations, i.e.,
\be
R(g_1)R(g_2)=\w(g_1,g_2)R(g_1g_2).\ee
Here $\w(g_1,g_2)\in U(1)$ is the factor set corresponding to the projective representation in question. Were $R$ a one dimensional representation, i.e., $R(g)=e^{i\phi(g)}$ then 
\be
\w(g_1,g_2)={e^{i\phi(g_1)}e^{i\phi(g_2)}\over e^{i\phi(g_1g_2)}}\times 1.\label{equiv}\ee
\Eq{equiv} is precisely the equivalence relation between $\w(g_1,g_2)$ and $1$, namely the linear representation, of $G$. This contradicts the assumption that $\w$ is the factor set of a projective representation. The same proof goes through if $G$ contains anti-unitary transformation. For example if $g_1$ is antiunitery then \Eq{equiv} is modified to \be
\w(g_1,g_2)={e^{i\phi(g_1)}e^{\rho(g_1)i\phi(g_2)}\over e^{i\phi(g_1g_2)}}\times 1.\label{equiv2}\ee where $\rho(g_1)$ is $(-) 1$ if $g_1$ is (anti-)unitary), which is the equivalence condition in the presence of anti-unitary group elements.\\

This theorem allows us to conclude that {\it the conformal spectrum for all critical points of SPT phase transitions possess states that transform as projective representation of the protection symmetry group.} In the following we argue that this also occurs for the ground states of certain conformal tower. Through the state - operator correspondence this means there are primary scaling operators that transform projectively under the protection symmetry group. \\

\begin{figure*}
\includegraphics[scale=.4,angle =0]{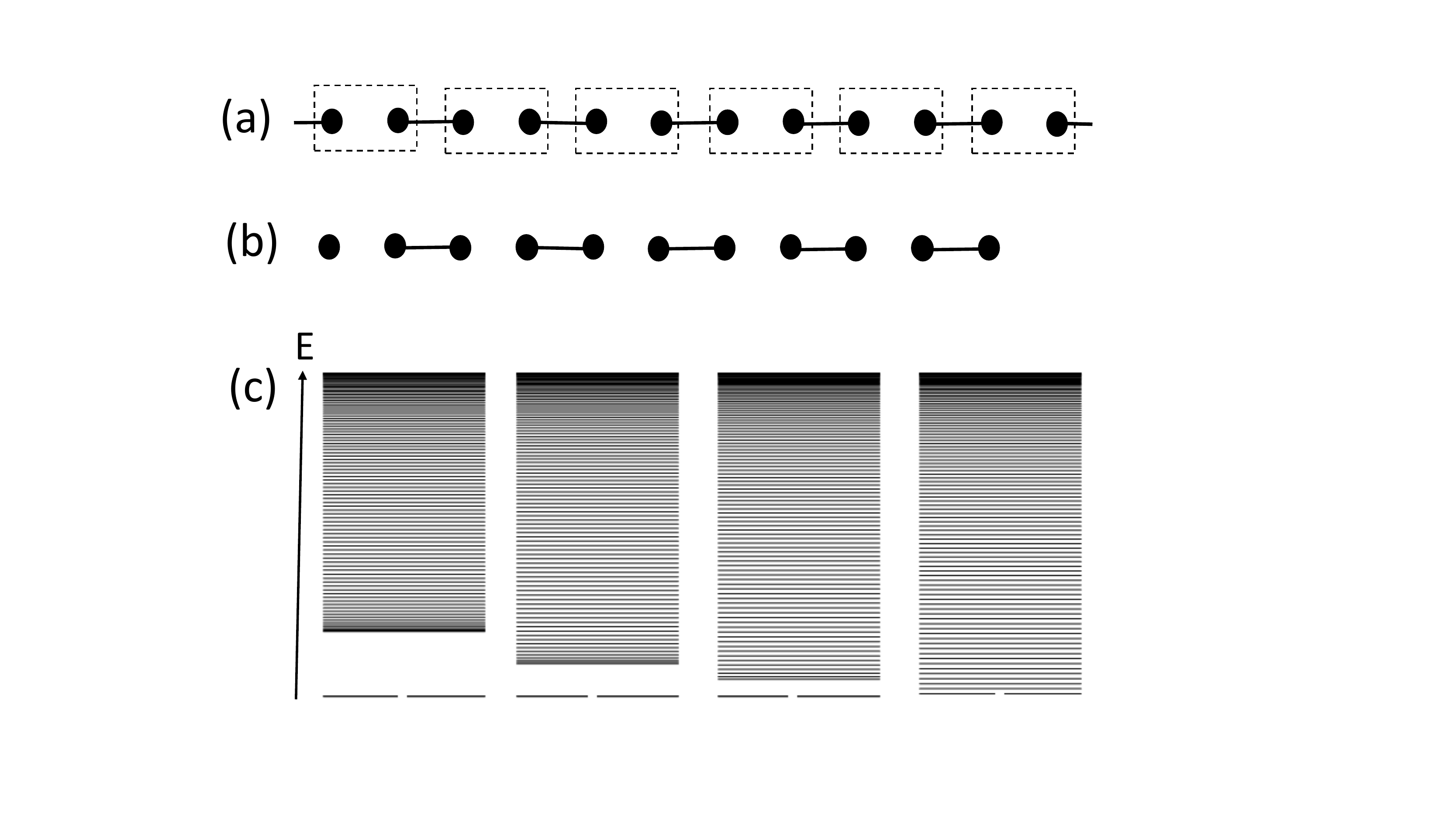}[h]
\caption{(a) A schematic representation of the entanglement pattern of 1D SPTs. Here each black dot represent degrees of freedom which transform projectively under the protection symmetry group. The dashed boxes represent the unit cell. The degrees of freedom in each unit cell transform linearly. (b) The boundary condition corresponding to removing one ``site''. There are multiply degenerate boundary zero energy excitations which transform projectively. (c) The excitation spectrum as the tuning parameter $\lambda$ approaches the critical value $\lambda_c$. Here the boundary states are assumed to be two-fold degenerate. The gap gradually closes as $\lambda\ra\lambda_c$ and the edge states become the ground states of a conformal tower.}
\label{entang}
\end{figure*}

To achieve the above goal we recall that in \Ref{Chen2014} it is shown that after proper local unitary transformation the fixed point ground state wavefunctions of 1D SPTs exhibit the entanglement
pattern shown in \Fig{entang}(a). Each black dot represents a group of degrees of freedom which carry a projective representation of the protection symmetry group and those in each dashed box transform linearly. In \Fig{entang}(b) the boundary condition corresponding to removing a ``site'' is imposed. Under such boundary condition the finite size spectrum as a function of the tuning parameter, $\lambda$,  is shown schematically in \Fig{entang}(c). As $\lambda\ra\lambda_c$ (left to right) the gap gradually closes, but the ground state remain degenerate and transform in the same way as the edge excitations. At $\lambda_c$ the edge excitations evolve into the ground state of a conformal tower. It is 
known that in an open conformal invariant system the energy spectrum consists of only the conformal towers in the holomorphic sector (rather than the sum of the holomorphic and antiholomorphic sector)\cite{Cardy2006}. Moreover the boundary conditions are represented by the insertion of appropriate operators at the infinite past and future in the strip geometry\cite{Cardy2006}. Therefore the above arguments imply that the boundary condition corresponding to the removal of a ``site'' amounts to inserting the primary scaling operator creating the delocalized edge excitation in the holomorphic sector. Note that the above arguments hold true for all continuous 1+1 D SPT phase transitions, hence has gone beyond the holographic theory in \Ref{Lokman2015}. In the following we give two examples where what's stated are explicitly realized. 
\\     

{\underline{$G=SO(3)$}}
The critical theory of the trivial to non-trivial SPT phase transition is the level 1 $SU(2)$ Wess-Zumino-Witten theory\cite{Affleck1987}. Such theory has 
the ``spinon'' primary scaling operators which transform as projective representation of $SO(3)$. The multiplicity and their holomorphic/anti holomorphic scaling dimensions ($h$ and $\bar{h}$) and the eigenvalue with respect to dilatation ($h+\bar{h}$)are given in Table \ref{so(3)}.The spinon states appear as the ground state of the open Heisenberg chain with an odd  (i.e., even-1) number of sites. \\
\begin{table*}[h]
\centering
\begin{tabular}{ |c|c|c|c|}
\hline
{Multiplicity}	&{$h$}	&$\bar{h}$	&$h+\bar{h}$\\
\hline
2					&1/4	&0			&1/4\\
\hline
2				&0		&1/4		&1/4\\
\hline
\end{tabular}
\caption{Quantum numbers associated with the spinon primary scaling operator in the conformal field theory (the $SU(2)_1$ WZW model\cite{Francesco2012}). The columns labeled by $h$ and $\bar{h}$ give the holomorphic and antiholomorphic scaling dimensions and $h+\bar{h}$ is the eigenvalue with respect to dilatation.}
\label{so(3)}
\end{table*}

{\underline{$G=Z_2\times Z_2$}} 
Again there are only two inequivalent classes 
of SPTs with this symmetry in 1D. In Ref.\cite{Lokman2015} an exact solvable Hamiltonian (which turns out to be the XX model) for the critical of phase transition was obtained. The primary scaling operator which transform projectively under $Z_2\times Z_2$ and its quantum numbers are given in Table \ref{z2z2}. Again by studying an open system with odd number of sites the ground states correspond to this scaling operator.\\ 
\begin{table*}[h]
\centering
\begin{tabular}{ |c|c|c|c|}
\hline
{Multiplicity}	&{$h$}	&$\bar{h}$	&$h+\bar{h}$\\
\hline
2		&9/32	&1/32		&5/16\\
\hline
2		&1/32	&9/32		&5/16\\
\hline
\end{tabular}
\caption{Quantum numbers associated with the primary scaling operator in the conformal field theory describing the $Z_2\times Z_2$ phase transition (the XX model\cite{Talstra1997}) which transform projectively under the protection symmetry group. The columns labeled by $h$ and $\bar{h}$ give the holomorphic and antiholomorphic scaling dimensions and $h+\bar{h}$ is the eigenvalue with respect to dilatation.} 
\label{z2z2}
\end{table*}

Now we return to the conformal field theories for Landau-like phase transitions in 1+1 D (either 2D classical or 1D quantum). By far the best studied such conformal field theories are the unitary ``minimal models''\cite{Belavin1984}. These conformal field theories possess a {\it finite} number of primary scaling operators and the operator product expansion is closed among them. Each of these theories is characterized by a parameter call ``central charge''. They are given by 
\be
c=1-{6\over m(m+1)}~~{\rm where~~} m=3,4,5,...\ee Hence for finite $m$ these central charges are less then 1.
For each fixed central charge (hence $m$) the holomorphic scaling dimension of the primary operators  are given by
\be
h_{r,s}={[r(m+1)-s m]^2-1\over 4m(m+1)}~~{\rm where~~}1\le s\le r\le m-1.\ee
The same expression holds for the antiholomorphic scaling dimensions.             
It can be shown easily that the closest spacing between such scaling dimensions are given by
\be
\Delta h_{min}={(1-c)\over 8}.\ee
Hence for $c<1$ all minimal models do not possess degenerate primary scaling operators required by the projective representations. Thus we conclude none of these best studied critical theories for Landau-type phase transitions can describe the critical point of bosonic SPT phase transitions.\\

The above results are consistent with the the examples of bosonic SPT phase transitions presented earlier. In addition it is also consistent with a new $Z_3\times Z_3$ protected SPT phase transition\footnote{Hongchen Jiang {\it et al} to be published.}. The following table summarize their central charge. \\
\begin{table*}[h]
\centering
\begin{tabular}{ |c|c|c|c|}
\hline
{Symmetry group}	&{$SO(3)$}	&$Z_2\times Z_2$	&$Z_3\times Z_3$\\
\hline
{Central charge}	&1	&1		&8/5\\
\hline
\end{tabular}
\caption{The central charge associated with three 1D SPT phase transitions. The result on the $Z_3\times Z_3$ transition by Hongchen Jiang {\it et al} is yet to be published. } 
\label{z3z3}
\end{table*}

For fermion SPT phase transitions $c<1$ is allowed. For example the critical point of the trivial to non-trivial phase transition of the Majorana fermion chain has central charge $c=1/2$. However for the fermion theories the statement that the edge excitation carries the projective representation of the protection symmetry group does not hold.\\

Many questions remain open concerning the SPT phase transitions. Here let's name a few. (1) From the holographic picture presented in \Ref{Lokman2015} we conclude that for $d>1$, the critical theory possesses fluctuating manifolds each has gapless excitations residing on it. We do not know a concrete example for such phase transition. (2) For D=1+1 which subset of the $c\ge 1$ conformal field theory can describe SPT phase transitions? (3) Is there always emergent continuous internal symmetry at the critical point of SPT phase transitions? (4) Can one exploit the holographic correspondence and learn something about the critical theory of SPT phase transition from the bulk gapped phase? Clearly much future studies are warranted for the understanding of these interesting phase transitions.\\

{{\bf Acknowledgement}} DHL would like to thank Martin Greiter, Ronny Thomale, Vera Schnells and Guang-Ming Zhang for useful discussions. This project was conceived while DHL was appointed the 2015 Wilhelm Wien Professor of the  University of W{\"u}rzburg. DHL was supported by the U.S. Department of Energy, Office of Science, Basic Energy Sciences, Materials Sciences and Engineering Division, grant DE-AC02-05CH11231.

\newpage
\bibliographystyle{ieeetr}
\bibliography{bibs}
\end{document}